\documentclass[a4paper]{jpconf}
\usepackage{graphicx}
\begin{document}
\title{Spectral synthesis of circumstellar disks - application to
  white dwarf debris disks} 

\author{S. D. H\"ugelmeyer$^1$, S. Dreizler$^1$, D. Homeier$^1$, and P. Hauschildt$^2$}

\address{$^1$Institut f\"ur Astrophysik, Georg-August-Universit\"at
  G\"ottingen, Friedrich-Hund-Platz~1, 37077~G\"ottingen, Germany}
\address{$^2$Hamburger Sternwarte, Gojenbergsweg 112, 21029 Hamburg,
    Germany}

\ead{shuegelm@astro.physik.uni-goettingen.de}

\begin{abstract} 
Gas and dust disks are common objects in the universe and can be found
around various objects, e.g. young stars, cataclysmic variables,
active galactic nuclei, or white dwarfs. The light that we receive
from disks provides us with clues about their composition,
temperature, and density. In order to better understand the physical
and chemical dynamics of these disks, self-consistent radiative
transfer simulations are inevitable. Therefore, we have developed a
1+1D radiative transfer code as an extension to the well-established
model atmosphere code \verb!PHOENIX!. We will show the potential of
the application of our code to model the spectra of white dwarf debris
disks.
\end{abstract}

\section{Introduction}

The detection of infrared excesses in association with the
metal rich white dwarfs GD~362 and G29-38 has introduced the picture
of debris disks around these stars. The high metal content observed in
these evolved stars' atmospheres requires a steady or recent source of
enriched matter, since the metals could only survive depletion by
gravitational settling for a very short time.  This source is believed
to be provided by accretion of dusty material from a disrupted solid
body, such as an asteroid. The IR excess is a strong indication for
the formation of an accretion disk. To understand the origin of the
dust, as well as to constrain the accretion rate and thus models for
the gravitational diffusion of heavy elements in the atmosphere,
comparison of observations with model spectra are necessary.

\section{Standard accretion model}

We adopt the standard accretion model for geometrically thin disks,
i.~e.~height $H$ $\ll$ radius $R$ (Shakura \& Syunyaev 1973). Matter
is assumed to rotate with Kepler velocity and viscous shear
decelerates inner and accelerates outer parts leading to accretion of
matter and outward transportation of angular momentum. Turbulent cells
smaller than the disk height $H$ are the proposed origin of kinematic
viscosity. The height averaged kinematic viscosity is usually
described by
\begin{equation} 
  \nu=\alpha c_s H \quad ,
\end{equation}
where $0 \le \alpha \le 1$ is the angular
momentum transfer efficiency.

\section{Model calculations}

We use a 1+1D disk ring structure approximation. The disk is divided
into rings and each ring is assumed to be plane-parallel and to have
constant physical properties from its inner to its outer
radius. Radiative transfer and disk structure are calculated
vertically from the disk's midplane to the top layer. Gas and dust are
in chemical equilibrium and dust is assumed not to settle in the disk.

\vspace{-.5in}
\begin{figure}[h]
  \begin{minipage}[b]{\textwidth}
    \begin{minipage}[c]{0.6\textwidth}
      \includegraphics[bb=54 360 558
      600,clip,height=0.65\textwidth,width=0.99\textwidth]{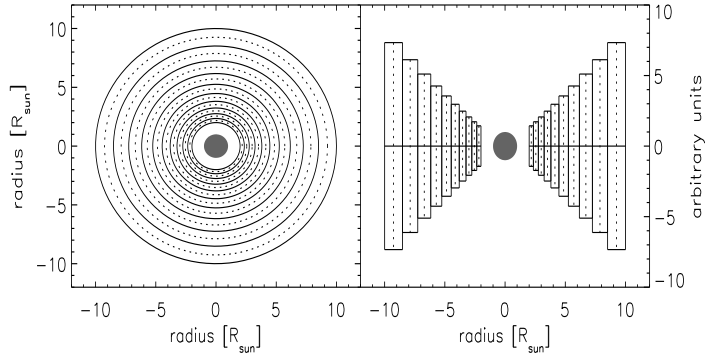}
    \end{minipage} \begin{minipage}[c]{0.35\textwidth}
      \caption{\label{fig:geo}Disk structure as adopted for our
        numerical calculations. Radial disk ring extensions are chosen
        so that the structure varies only little over ring width. The
        disk is thin ($H \ll R$) so that vertical and radial structure
        can be separated.}
    \end{minipage}
  \end{minipage}  
\end{figure}

Basic program input parameters are the radius $R_\star$ and mass
$M_\star$ of the central star, the distance $R$ between star and disk
ring, the mass accretion rate $\dot{M}$ in the disk, and the Reynolds
number $Re$ as a measure for the mean kinematic viscosity
$\nu=\sqrt{G M_\star R}/Re$.

The iterative calculation of a disk ring model atmosphere starts by
either constructing a grey start model for disks (after Hubeny 1990)
or using a given \verb!PHOENIX! model. We have further adopted the
following conservation and constraint equations:\\

\begin{figure}[ht!]
    \centering
    \includegraphics[bb=221 213 1043 600,clip,width=0.58\textwidth]{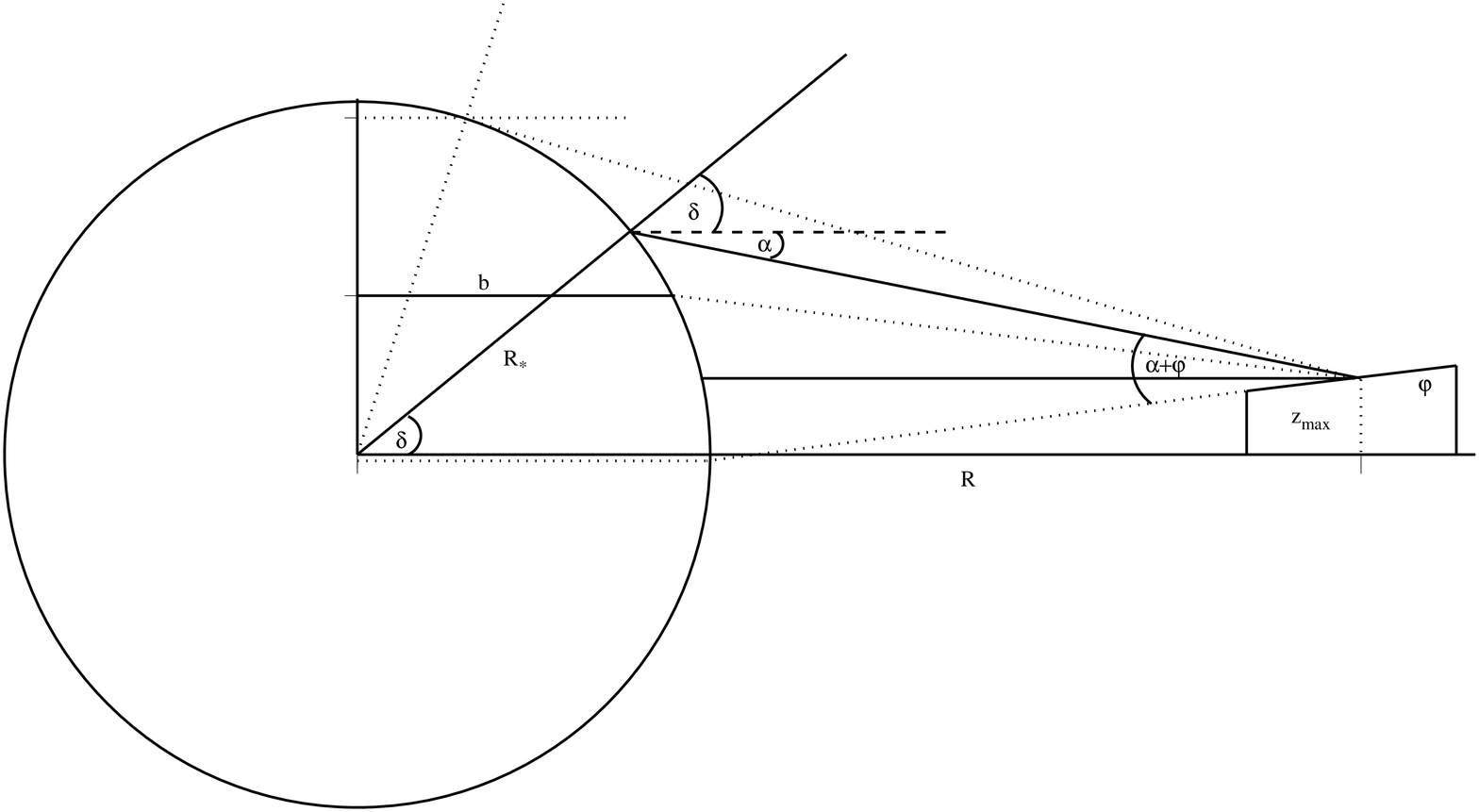}
    \includegraphics[bb=-100 236 473 473,clip,width=0.34\textwidth]{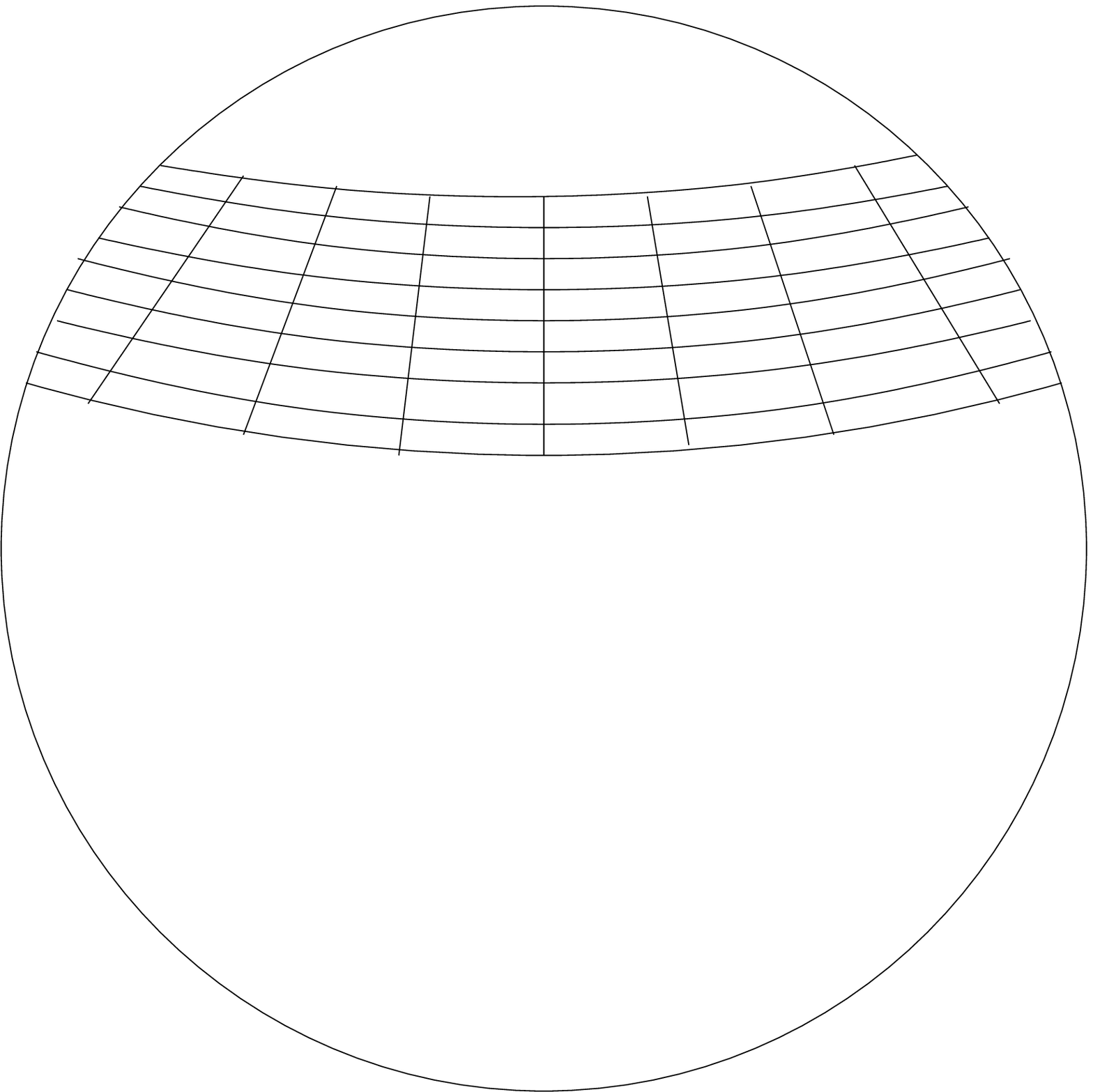}    
    \caption{\textit{Left}: Star -- disk irradiation
      geometry. \textit{Right}: The star's surface fraction is
      subdivided in sections and the irradiation flux is calculated
      for each ray considering limb darkening.}
    \label{fig:irr}
  \end{figure}

\noindent \textbf{Hydrostatic equilibrium:} This equation has to be
used in a form to account for the varying surface gravity, i.~e.
\begin{equation}
  \frac{dP}{dm}=\frac{GM_\star}{R^3} z \quad .
\end{equation}
\noindent \textbf{Energy conservation:} Unlike ordinary stellar
  atmospheres, mechanical energy is released in disk atmospheres in
  every layer. Therefore, the flux is not conserved and we have to
  consider energy equilibrium of the form
\begin{equation}
  E_{\rm mech}=E_{\rm rad} + E_{\rm conv} \quad .
\end{equation}
\noindent \textbf{Irradiation:} Also irradiation from the
  central star can be considered. As input source serves one or a
  combination of black body spectra of given effective temperatures. A
  \verb!PHOENIX! spectrum can also be used as input. Discrete rays
  that receive radiation from the star are determined and assigned a
  fraction of the flux incident to the disk. The irradiation geometry
  is shown in Figure~\ref{fig:irr}.

\section{Results}

We have calculated a simple disk model consisting of three disk
rings. The rings are in hydrostatic equilibrium and the mass accretion
rate is set according to Becklin et al. (2005) to $\dot{M}_\star
= 10^{11}~{\rm g~s^{-1}} \approx 2\cdot 10^{-15}~M_\odot~{\rm
yr}^{-1}$. This small accretion rate leads to small surface densities
($\doteq$ disk mass) and the disk rings are optically thin in the
continuum. The vertical temperature distribution is almost constant
and only ``artificially'' brought to a value reasonably in agreement
with the blackbody temperature fit to the infrared excess by setting
an unrealistically high viscosity value.

In order to estimate the mean dust grain size, we have performed
various computations varying this parameter. This allows us to get an
idea of the influence on the strength of the silicate feature at
$10~\mu{\rm m}$ and the $13~\mu{\rm m}$ emission. We have varied the
mean dust grain size between $0.01~\mu{\rm m}$ and $10~\mu{\rm
m}$. The resulting spectra of these simulations are shown in
Figure~3. For our refined models of GD~362 we have chosen a mean dust
grain size of $0.01~\mu{\rm m}$.

\begin{figure}[ht!]
  \centering
  \begin{minipage}[b]{\textwidth}
    \begin{minipage}[c]{0.6\textwidth}
      \includegraphics[width=0.99\textwidth]{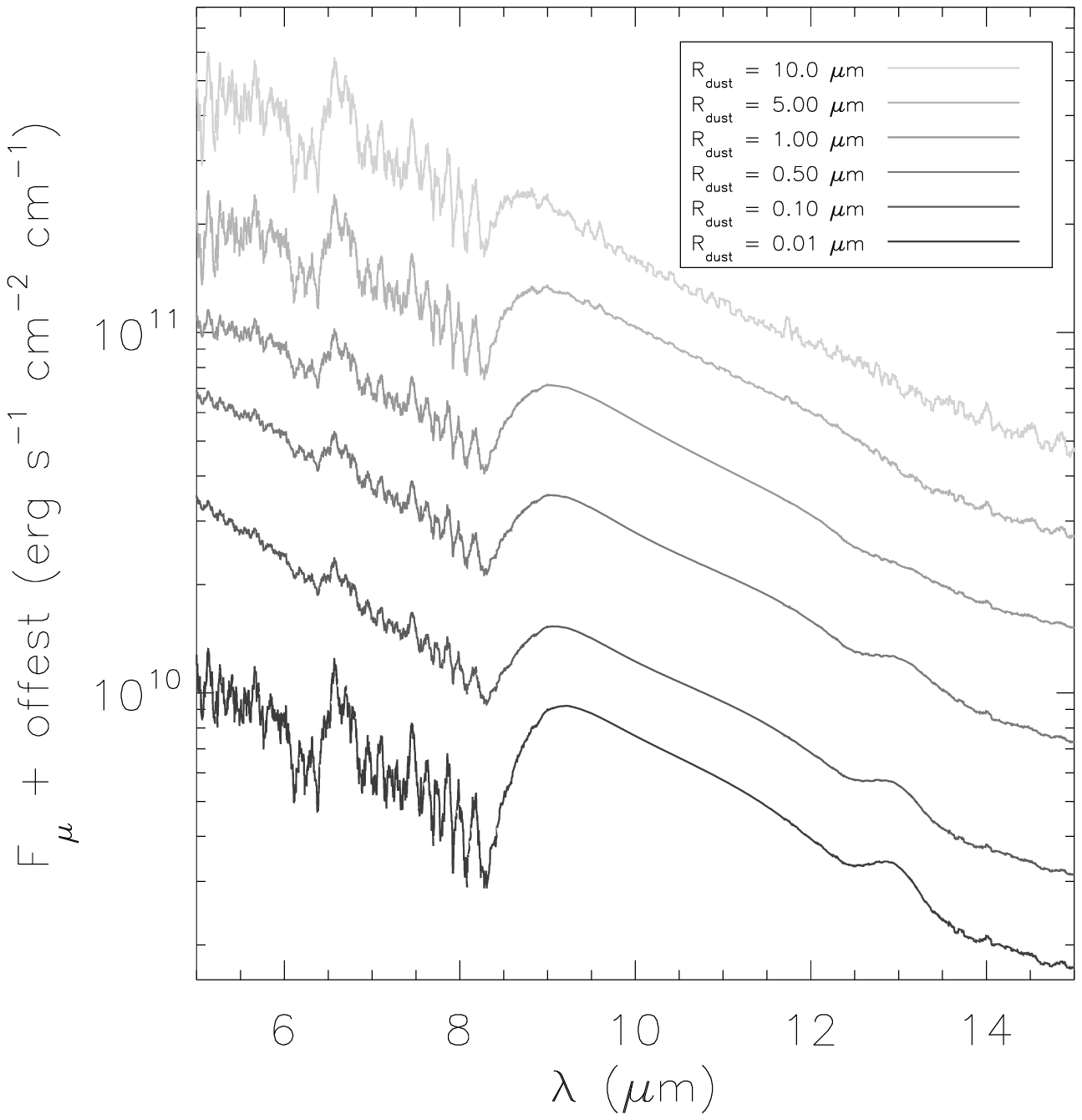}
    \end{minipage} \begin{minipage}[c]{0.35\textwidth}
      \vspace{-1.5in}
      \caption{\label{fig:gs}Spectra in the range of the
        characteristic $10~\mu{\rm m}$ silicate feature for different
        mean dust grain sizes. For these calculations we have assumed
        $M_\star=0.6~M_\odot$, $R_\star=0.01~R_\odot$,
        $R=100~R_\star$, and $\dot{M}=10^{-11}~M_\odot~{\rm
          yr}^{-1}$. For this setup, the smallest grain size produces
        the most prominent $13~\mu{\rm m}$ feature which is also
        clearly visible in the {\it Spitzer}
        spectrum} 
    \end{minipage}
  \end{minipage}
\end{figure}

Since the disk slabs are optically thin, the continuum flux is in a
reasonable approximation linear with the optical depth. Therefore, the
fitting of the models' continuum to the {\it Spitzer} spectrum can be
adjusted by varying the mass accretion rate. We simply used a fudge
factor to account for the missing flux.

\begin{figure}[ht!]
  \centering
  \includegraphics[width=\textwidth]{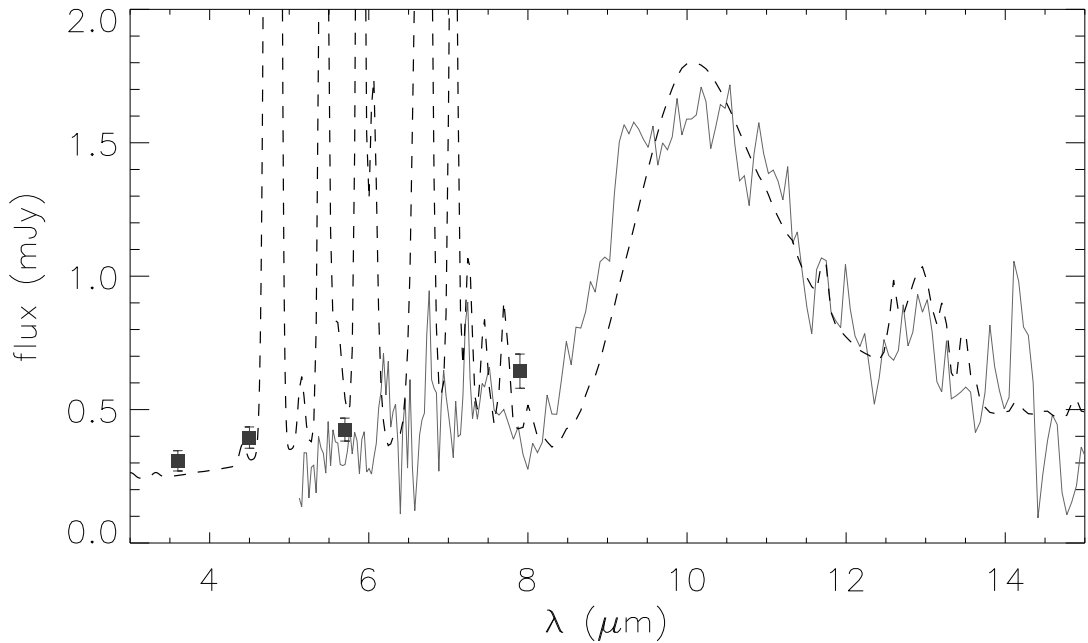}
  \caption{\label{fig:sp}{\it Spitzer} IRS spectrum (grey line;
    Farihi, private communication) and photometry (squares with error
    bars; Jura et al. 2007) of GD~362. Our model fit is plotted as
    dashed line (disk+star) and reproduces the $10~\mu{\rm m}$
    silicate and the $13~\mu{\rm m}$ feature fairly well.}
\end{figure}

In Figure~\ref{fig:sp} we show the co-added models folded to the
resolution of the instrument. The innermost ring is directly
irradiated and heated up by the white dwarf and is assumed to be
isothermal ($700~{\rm K}$), thus shielding the disk parts behind
it. The inner ring extends from $9.3 - 10~R_\star$. The following
three rings go from $10 - 70~R_\star$ and have almost isothermal
temperature structures between $670~{\rm K}$ and $500~{\rm K}$.

\section{Discussion}

We have shown that our disk model atmosphere code is in principal
applicable to the modelling of white dwarf debris disks. Our models
calculated for the parameters of GD~362 show a reasonable fit to IR
observations of the objects. Masses and accretion rates of WD debris
disks are however small and therefore make the model calculations
challenging.

\section*{Acknowledgments} SDH is supported by a
  scholarship of the DFG Graduiertenkolleg 1351 ``Extrasolar Planets
  and their Host Stars''

\section*{References}

\begin{thereferences}
\item {Becklin}, E.~E., {Farihi}, J., {Jura}, et al. M. 2005 {\sl ApJ}
  {\bf 632} L119
\item {Hubeny}, I. 1990 {\sl ApJ} {\bf 351} 632
\item {Jura}, M., {Farihi}, J., and {Zuckerman}, B. 2007 {\sl ApJ}
  {\bf 663} 1285
\item {Shakura}, N.~I. \& {Syunyaev}, R.~A. 1973 {\sl A\&A} {\bf 24} 337
\end{thereferences}


\end{document}